# Girth of a Planar Digraph with Real Edge Weights in $O(n \log^3 n)$ Time

Christian Wulff-Nilsen [*]

## Abstract

The girth of a graph is the length of its shortest cycle. We give an algorithm that computes in $O(n \log^3 n)$ time and $O(n)$ space the (weighted) girth of an $n$-vertex planar digraph with arbitrary real edge weights. This is an improvement of a previous time bound of $O(n^{3/2})$, a bound which was only valid for non-negative edge-weights. Our algorithm can be modified to output a shortest cycle within the same time and space bounds if such a cycle exists.

## 1 Introduction

The *girth* of an unweighted graph is the length of a shortest cycle in the graph or $\infty$ if the graph is acyclic. It is a well-studied graph characteristic and has been shown to be related to numerous other properties of graphs, including vertex degree, diameter, connectivity, maximum genus, and vertex colourings [2, 4]. For instance, it is easy to see that for a graph containing a cycle, its girth is at most twice its diameter plus one.

The problem of computing the girth of a graph has received some attention. An $O(mn)$ time algorithm is known where $m$ and $n$ are the number of edges and vertices, respectively [6]. Finding the length of a shortest cycle of even length can be done in $O(n^2)$ time [11].

In this paper, we focus on the problem of computing the girth of planar graphs. For this class of graphs, faster algorithms are known. A linear time algorithm is presented in [3] but it only applies when the graph has bounded girth. For general planar graphs, Djidjev gave an $O(n^{5/4} \log n)$ time algorithm. This can be improved to $O(n \log^2 n)$ time by applying the minimum cut algorithm of [1] to the dual graph. Recently, it was shown how to find the girth in $O(n \log n)$ time [10].

[*]Department of Computer Science, University of Copenhagen, koolooz@diku.dk, http://www.diku.dk/~koolooz/

All these results for planar graphs assume that the graph is undirected. For planar digraphs, Weimann and Yuster [10] gave an $O(n^{3/2})$ time algorithm and they asked whether a faster algorithm exists.

The girth of a graph is defined for unweighted graphs but this definition immediately extends to the case where edges have real weight. Of the above algorithms for planar graphs, only the $O(n \log^2 n)$ time algorithm in [1] and the $O(n^{3/2})$ time algorithm in [10] can handle weighted graphs and only if all edge weights are non-negative. The $O(n \log^2 n)$ time algorithm only applies when the graph is undirected.

We consider the most general version of the problem for planar graphs. We show how to find the girth of an $n$-vertex planar digraph with arbitrary real edge weights (non-negative as well as negative) in $O(n \log^3 n)$ time and $O(n)$ space. In particular, we answer the question by Weimann and Yuster [10] of whether an $o(n^{3/2})$ time algorithm exists for computing the girth of a planar digraph. Our algorithm can output a shortest cycle within the same time and space bounds, assuming such a cycle exists.

The organization of the paper is as follows. In Section 2, we introduce some basic definitions and notation. We give our algorithm for computing the girth of a planar digraph in Section 3 and in Section 4, we show how to extend this algorithm to output a shortest cycle. Finally, we make some concluding remarks in Section 5.

## 2 Definitions and Notation

For a graph $H$, we let $V_H$ and $E_H$ denote its vertex and edge set, respectively. Let $G = (V, E)$ be a digraph with real edge weights defined by weight function $w : E \to \mathbb{R}$. For vertices $u, v \in V$, we let $d_G(u, v)$ denote the length of a shortest path in $G$ from $u$ to $v$ w.r.t. $w$ (we omit $w$ in the notation but this should not cause any confusion). If no path from $u$ to $v$ exists, we define $d_G(u, v) = \infty$ and if there is a path from $u$ to $v$ containing a negative-weight cycle, $d_G(u, v) = -\infty$.

The *(weighted) girth* of $G$ is the length of a shortest cycle in $G$ w.r.t. $w$. If $G$ is acyclic, we define its girth to be $\infty$. If $G$ contains a negative-weight cycle $C$, cycles of arbitrarily large negative weight can be obtained by traversing $C$ sufficiently many times so in this case, we define the girth of $G$ to be $-\infty$.

## 3 Computing the Girth

In the following, let $G = (V, E)$ be an $n$-vertex planar digraph with real edge weights defined by weight function $w : E \to \mathbb{R}$. In this section, we show how to compute the girth of $G$ in $O(n \log^3 n)$ time and $O(n)$ space.

We may assume that $G$ contains no negative-weight cycles since the algorithm in [5] can detect such cycles within our time and space bounds. If a negative-weight cycle is present, our algorithm outputs $-\infty$ as the girth of $G$.

We will assume that $G$ is triangulated with pairs of oppositely directed edges. If it is not, this can be achieved by adding edges of sufficiently high weight $W$ so that finite shortest path distances in $G$ will not be affected. We define
$$W = 1 + 2 \sum_{e \in E'} |w(e)|,$$
where $E'$ is the set of edges in the original graph. This way, we avoid dealing with infinite shortest path distances. And we can still detect the case where the girth is $\infty$ in the original graph since this holds if and only if the girth of the triangulated graph is at least $1 + \sum_{e \in E'} |w(e)|$.

We will identify $G$ with a fixed plane embedding of the graph.

Next, we apply the linear time cycle separator theorem of Miller [9] to $G$. This gives a simple cycle $C$ in $G$ (ignoring edge orientations) containing $O(\sqrt{n})$ vertices such that at most $2n/3$ vertices of $G$ are in the closed bounded region $R_1$ resp. closed unbounded region $R_2$ of the plane defined by $C$. Let $G_1$ resp. $G_2$ be the subgraph of $G$ containing the set of vertices and edges of $G$ in $R_1$ resp. $R_2$. If an edge of $G$ belongs to both $R_1$ and $R_2$, i.e., to $C$, then we only add it to one of the two subgraphs, say $G_1$. This ensures that $G_1$ and $G_2$ are edge-disjoint.

We recursively compute the girth $g_1$ of $G_1$ and the girth $g_2$ of $G_2$. Let $g$ denote the length of a shortest cycle in $G$ that contains at least two vertices of $C$. Any simple cycle in $G$ that is neither fully contained in $G_1$ nor in $G_2$ must contain at least two vertices of $C$. Thus, $\min\{g_1, g_2, g\}$ is the girth of $G$ so let us consider the problem of computing $g$.

Let $H$ be the complete digraph with vertex set $V_H = V_C$ and with weight function $w_H : E_H \to \mathbb{R}$ defined by $w_H(u,v) = \min\{d_{G_1}(u,v), d_{G_2}(u,v)\}$ for all distinct $u, v \in V_H$.

For any $u, v \in V_H$, a shortest path from $u$ to $v$ in $G$ can be decomposed into subpaths each of which has the property that it is a shortest path in either $G_1$ or in $G_2$ with both its endpoints on $C$. It follows that $d_H(u,v) = d_G(u,v)$.

We apply the algorithm in [7] to compute $d_{G_1}(u,v)$ and $d_{G_2}(u,v)$ for all $u, v \in V_C$ using a total of $O(n \log^2 n)$ time over all recursion levels. Hence, we obtain $H$ and its edge weights in this amount of time over all recursion levels.

Any shortest cycle in $G$ containing at least two vertices of $C$ can be decomposed into subpaths each having the property above. Hence, such a cycle has the same length as a shortest cycle in $H$ so $H$ has girth $g$.

What remains therefore is to find the length of a shortest cycle in $H$. We will show how to do this in $O(n \log^2 n)$ time. Since there are $O(\log n)$ recursion levels, this will imply that the girth of $G$ can be obtained in $O(n \log^3 n)$ time.

Let $u_1, \ldots, u_m$ be the vertices of $C$. To compute the girth $g$ of $H$, we will compute, for $i = 1, \ldots, m$, the length $g_i$ of a shortest cycle in $H$ containing $u_i$. Then it is clear that we can obtain the girth of $H$ as $g = \min\{g_1, \ldots, g_m\}$.

We first reduce our problem to one where all edge weights are non-negative. This is done as follows. We compute single source shortest path distances with source, say, $u_1$, using the $O(n \log^2 n)$ time Bellman-Ford variant of [5] (in fact, it can be done in only $O(n\alpha(n))$ time with ideas from [8] but this will not improve our bound).

Then in $O(|E_H|) = O(n)$ additional time, we can obtain the *reduced cost* $w_H^+(e)$ of each edge $e = (u_i, u_j)$ of $H$ (w.r.t. $w_H$):

$$w_H^+(e) = d_H(u_1, u_i) + w_H(e) - d_H(u_1, u_j).$$

By the triangle inequality, $w_H^+ \geq 0$ and it is easy to see that for any cycle in $H$, its length w.r.t. $w_H$ is identical to its length w.r.t. $w_H^+$. This gives us the desired reduction.

For each pair of vertices $u_i$ and $u_j$ in $H$, we let $d_H^+(u_i, u_j)$ denote the shortest path distance from $u_i$ to $u_j$ in $H$ w.r.t. $w_H^+$.

Now, let us consider the problem of computing $g_i$ for some $i$. Since $g_i = \min\{d_H^+(u_i, u_j) + w_H^+(u_j, u_i) | j = 1, \ldots, m, j \neq i\}$, we can obtain this

value in $O(m) = O(\sqrt{n})$ time in addition to the time for computing single source shortest path distances from $u_i$ w.r.t. $w_H^+$. And since $w_H^+ \geq 0$, we can apply the Dijkstra variant of [5] to compute these distances in $O(m \log^2 n) = O(\sqrt{n} \log^2 n)$ time. Over all $i$, this is $O(n \log^2 n)$, as desired.

It follows from the above that we can find the girth of $H$ in $O(n \log^2 n)$ time and we can conclude this section with the following result.

**Theorem 1.** *The girth of an n-vertex planar digraph with real edge weights can be computed in $O(n \log^3 n)$ time and $O(n)$ space.*

*Proof.* We have shown the time bound above and since the algorithms in [5, 7, 8, 9] all require $O(n)$ space, this bound also holds for our algorithm. □

## 4 Finding a Shortest Cycle

We now show how to extend the algorithm of the previous section to compute a shortest cycle in $G$ within the same time and space bounds. We may assume that $G$ contains no negative-weight cycles since otherwise, a shortest cycle does not exist.

With the definitions above, it suffices to find in $O(n \log^2 n)$ time and $O(n)$ space a shortest cycle in $G$ containing at least two vertices of $C$. As already noted, such a cycle corresponds to a shortest cycle in $H$. The algorithm above finds vertices $u_i$ and $u_j$ of $C$ such that a shortest path $P_{ij}$ in $H$ from $u_i$ to $u_j$ followed by the edge $(u_j, u_i)$ is a shortest cycle $C'$ in $H$. The edges on $P_{ij}$ can be found by traversing the shortest path tree in $H$ rooted at $u_i$. Since this shortest path tree has already been computed, we can thus find all edges of $C'$ within the required time and space bounds.

What remains is to replace each edge $e$ of $C'$ by a simple shortest path $P_e$ in $G$ between its endpoints. We will show how to do this in $O(n \log n)$ additional time.

To obtain these paths, we need to take a closer look at the multiple-source shortest path algorithm of Klein [7] which we applied to find the weights of edges of $H$. For $G_i$, $i = 1, 2$, his algorithm maintains a dynamic tree data structure, which is initally a shortest path tree in $G_i$ rooted at, say, $u_1$, then at $u_2$, and so on. To obtain the weights of edges of $H$, this data structure is repeatedly queried, first for the distance in $G_i$ from $u_1$ to all other vertices of $C$, then from $u_2$, et cetera.

Now, we also need the actual paths in the shortest path trees corresponding to these distances. Once the edges of $C'$ have been identified, we can run Klein's algorithm again. During the course of this second run of the algorithm, we can query the dynamic tree data structure to obtain the paths corresponding to edges of $C'$. Querying for a single vertex on a path takes $O(\log n)$ time so the total time is $O(k \log n)$, where $k$ is the total number of vertices (with repetitions) over all paths.

For this strategy to work, $k$ should not be too big. We will ensure this by modifying Dijkstra without increasing its time and space bounds such that $C'$ will have the least number of edges among all shortest cycles in $H$. We then show that none of the paths defining edges of $C'$ will share any edges of $G$, implying that $k = O(n)$ and hence that a shortest cycle in $G$ can be found in $O(n \log^3 n + k \log n) = O(n \log^3 n)$ time and $O(n)$ space.

To output a shortest cycle in $H$ with the minimum number of edges, we modify the Dijkstra algorithm in [5]. We will not go through all the details but assume that the reader is familiar with the paper.

During the course of that algorithm, we maintain not only shortest path distances w.r.t. $w_H^+$ but also w.r.t. the unit weight function. Now, whenever there is a tie between which vertex to pick next from the heap, we pick one for which the number of edges on a shortest path from the root of the partially built shortest path tree to $u_i$ is minimized. This can be incorporated into the algorithm without increasing its time and space bounds by lexicographically ordering vertices, first according to weighted and then according to unweighted distances from the root of the shortest path tree.

It remains to show that if $C'$ is a shortest cycle in $H$ and it is picked such that it has the minimum number of edges then none of the shortest paths in $G$ corresponding to edges of $C'$ share edges.

So let $P_1$ and $P_2$ be shortest paths in $G$ corresponding to distinct edges $e_1 = (u_i, u_j)$ and $e_2 = (u_{i'}, u_{j'})$ in $H$, respectively. Assume for the sake of contradiction that $e = (u, v)$ is an edge shared by $P_1$ and $P_2$. Let $P$ be the subpath of $C'$ starting in $u_j$ and ending in $u_{i'}$. Without increasing the length of $C'$, the subpath $P_1 P P_2$ of $C'$ can be replaced by the path $P'$ defined as the prefix of $P_1$ ending in $v$ followed by the suffix of $P_2$ starting in $v$. Since $G_1$ and $G_2$ are edge-disjoint, either $P_1$ and $P_2$ both belong to $G_1$ or both belong to $G_2$. Hence, $P'$ is a path in either $G_1$ or in $G_2$ so we can replace $e_1$ and $e_2$ by the edge $(u_i, u_{j'})$ in $H$. But this reduces the number of edges of $C'$ without increasing its length, contradicting the choice of the cycle.

It follows that none of the shortest paths in $G$ corresponding to edges of $C'$ share edges. By the above, this suffices to show the following.

**Theorem 2.** *A shortest cycle in an $n$-vertex planar digraph with real edge weights can be computed in $O(n \log^3 n)$ time and $O(n)$ space, assuming such a cycle exists.*

If $G$ has girth $-\infty$, a shortest cycle does not exist. But we can still output a negative-weight cycle within our time and space bounds by applying the algorithm in [5].

## 5 Concluding Remarks

We showed how to compute the girth of an $n$-vertex planar digraph with real edge weights in $O(n \log^3 n)$ time and $O(n)$ space. This is a significant improvement over the previous best bound of $O(n^{3/2})$ which only applied to planar digraphs with non-negative edge weights. We also showed how to output a shortest cycle without an increase in time or space, assuming such a cycle exists.

In [5], it is suggested that the results of that paper can be generalized to the class of bounded genus graphs. If this is true, we believe that a generalization of our algorithm to this class is also achievable.